\documentclass[11pt,twoside]{article}
\usepackage{./asp2014}

\aspSuppressVolSlug
\resetcounters

\begin{document}

\title{AGB Variables in Local Group Dwarf Irregular Galaxies}
\author{Patricia A. Whitelock,$^{1,2}$ 
\affil{$^1$ South African Astronomical Observatory, P.O.Box 9, 7935
           Observatory, South Africa; \email{paw@saao.ac.za}}
\affil{$^2$Department of Astronomy,
           University of Cape Town, 7701 Rondebosch, South Africa.}}

\paperauthor{Patricia A. Whitelock}{paw@saao.ac.za}{0000-0002-4678-4432}{SAAO}{UCT}{Cape Town}{Western Cape}{}{South Africa}

\begin{abstract} 
The near and mid-infrared characteristics of large amplitude, Mira, variables in Local Group dwarf irregular galaxies (LMC, NGC 6822, IC 1613, Sgr dIG) are described. Two aspects of  these variables are discussed. First, the short period ($P \lesssim 420$ days) Miras are potentially powerful distance indicators, provided that they have low circumstellar extinction, or can be corrected for extinction. These are the descendants of relatively low mass stars. 
Secondly, the longer period stars, many of which undergo hot bottom burning, are poorly understood. These provide new insight into the evolution of intermediate mass stars during the high mass-loss phases, but their use as distance indicators depends on a much firmer understanding of their evolution.

The change in slope of the $K$  period luminosity relation for O-rich stars that is seen around 400 to 420 days in the LMC is due to the onset of hot bottom burning. It will  be sensitive to metallicity and should therefore be expected at different periods in populations with significant differences from the LMC. 

The [4.5] period-luminosity relation splits into two approximately parallel sequences. The fainter one fits stars where the mid-infrared flux originates from the stellar photosphere, while the brighter one fits observations dominated by the circumstellar shell.

\end{abstract}

\section{Introduction}
I will focus on large amplitude AGB variables (Miras) and their importance both as distance indicators and as probes of stellar populations. A good deal of the discussion below is about the LMC, because of the availability of multi-epoch infrared data for this galaxy and because the on-going OGLE surveys \citep{Soszynski2009} have provided extraordinarily detailed insights into the variability of its AGB stars.   These detailed studies help interpret what we find in the more distant dwarf irregulars for which the observations are still rather sparse.  

Most of the $JHK_S$\footnote{\citet{Carpenter2001} provides a conversion between the SAAO $K$ mag \citep{Carter1990} and the 2MASS $K_S$ mag \citep{Skrutskie2006}; measures on both systems are mentioned in this review.}  observations discussed below were made with the Japanese-South African  1.4m InfraRed Survey Facility (IRSF), and give a tantalising view into what will be possible in more distant galaxies with much bigger telescopes.

As observing at infrared wavelengths becomes easier Miras will rival Cepheids as distance indicators \citep{Whitelock2014}. 
They are fundamental pulsators and  easily identified, and although their amplitudes are smaller in the infrared than at shorter wavelengths many stars still vary by more than a magnitude.
They are cool so we assume their spectra are dominated by molecular absorption. Depending on their mass and metallicity their atmospheres are either C- or O-rich and the molecules indicate which, but we don't always have spectra, particularly for faint sources. 

Pulsation periods are mostly between 100 and 500 days although there are examples up to and over 2000 days. They are very luminous, both bolometrically and in the infrared. So these will be the stars that you resolve in old and intermediate age populations, if you can resolve anything at all. Their evolution is still not well understood and details of the mass-loss process remain enigmatic despite significant progress on the theoretical side \citep[e.g.,][]{Whitelock2016}. They are responsible for some of the important galactic chemical enrichment, including carbon, lithium and s-process elements. As we find more of these stars in short-lived evolutionary phases in galaxies for which we know the distance, we should be able to develop our understanding of the late stages of stellar evolution.

Studies of Miras in Galactic Globular and Magellanic Cloud clusters, together with the kinematics of Galactic Miras indicate that the pulsation period of a Mira is a function of its initial mass \citep[e.g.,][and references therein]{Feast2009}, and that very little evolution of period occurs once a star starts pulsating in the fundamental mode. Alternatively the period evolution takes place extremely rapidly.

It has become common practice to distinguish Miras with very red colours from the others and to call them extreme-AGB (x-AGB) stars. Many of these stars were unknown prior to the surveys with Spitzer, some were called Miras, while others, particularly those found in the Galaxy were called OH/IR stars. In the following the term Mira is used for all large amplitude periodic AGB variables, independent of their colour or chemistry.

\section{Large Magellanic Cloud}
 The infrared period luminosity (PL) relation for Mira variables was first calibrated using the Magellanic Clouds \citep[e.g.,][]{Feast1989,Hughes1990}  and later shown to be just one of the PL relations followed by AGB stars  \citep[e.g.,][]{Ita2004}.  These various parallel PL relations could be understood as the result of different modes of pulsation, with the large amplitude Miras pulsating in the fundamental mode \citep{Wood1999, Wood2015}. 
 
\articlefiguretwo{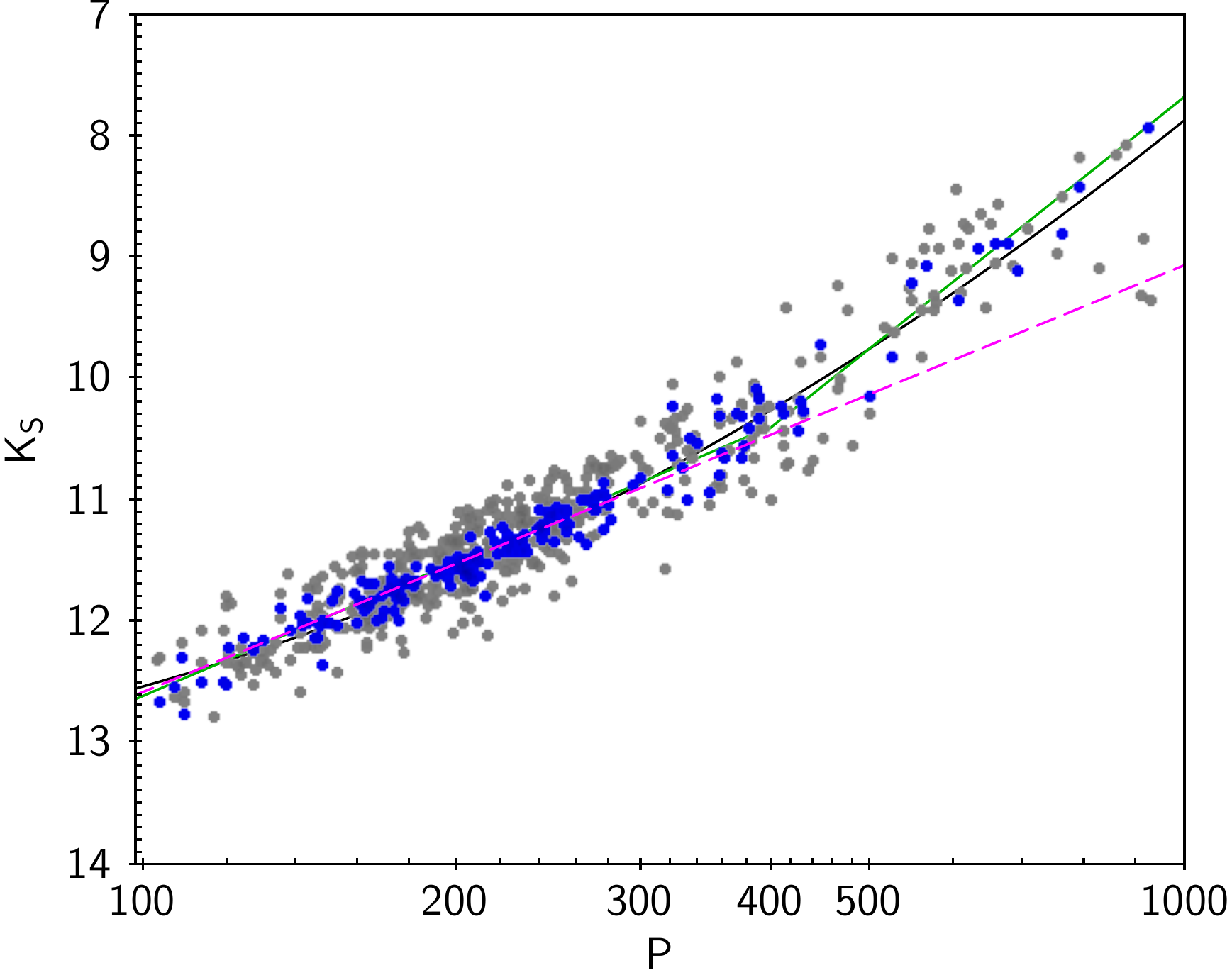}{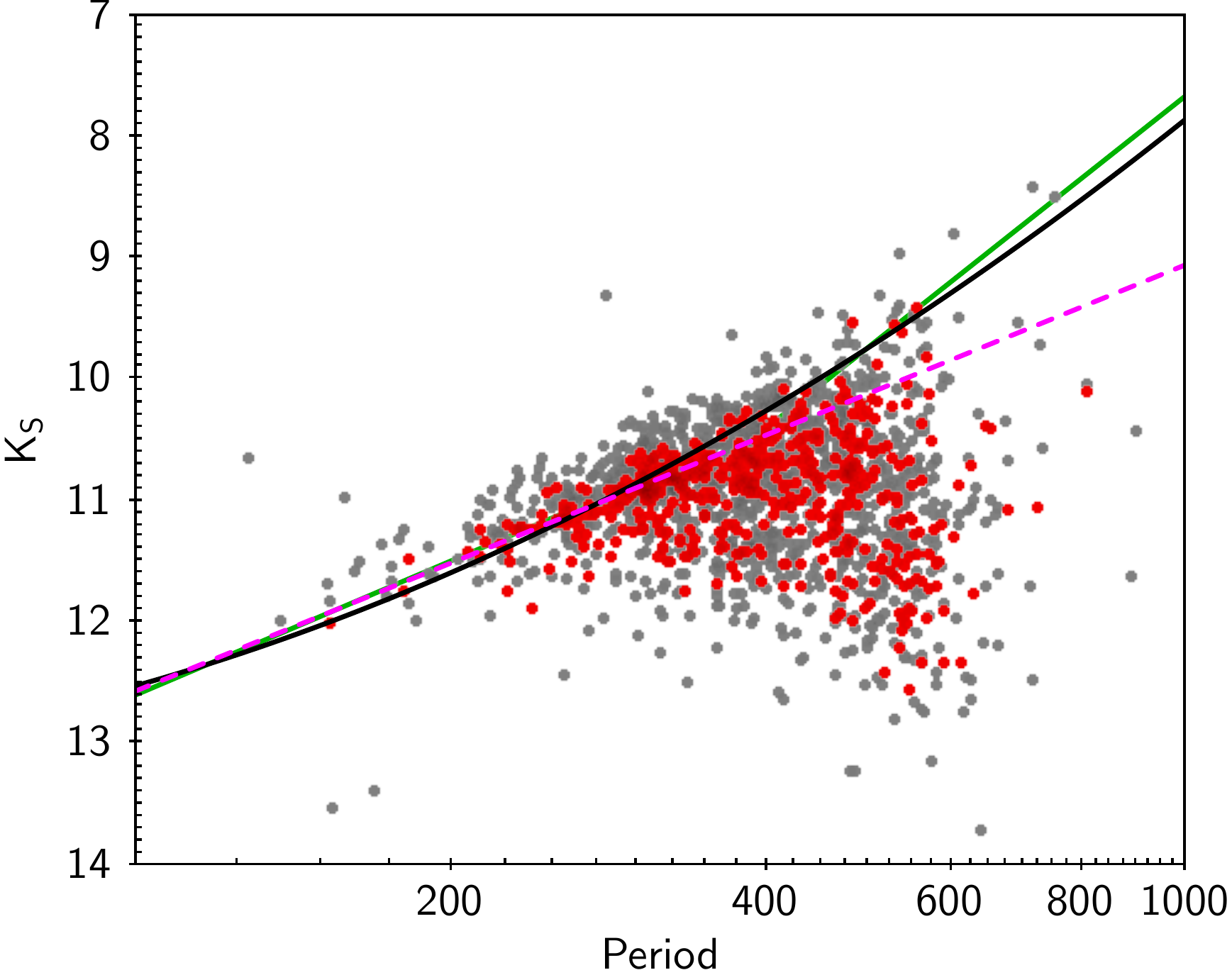}{LMC_PL}{PLRs for LMC Miras from 2MASS  (grey) and \citet{Yuan2017b} (blue/red); \emph{left:} O-rich stars \emph{right:} C-rich stars. The lines are the PL relations from \citet{Yuan2017b}(green), \citet{Ita2011}(black) and \citet{Whitelock2008}(magenta). }

\citet{Whitelock2008} discussed the luminosities of Galactic Miras and showed that they followed the same PL relation at $K$ as did Magellanic Cloud stars. This indicated that metallicity effects on the PL relation must be small, at least at $K$. Gaia will of course allow a much more detailed calibration to be made for Galactic AGB stars. 

It has been known since the work of \citet{Feast1989} that there are O-rich Miras with periods over 400 days that were more luminous than the $K$ PL relation for shorter period stars; \citet{Whitelock2003} suggested that these were hot bottom burning (HBB) stars.  It is also well known that there are significant numbers of Miras, mainly C-rich, that fall below the $K$ PL relation because their mass-loss rates are high and circumstellar extinction reduces their $K$ luminosity.

\citet{Ita2011} fitted two linear PL relations to the $K_S$-band photometry of O-rich LMC Miras with a break point at 400 days. These showed a good deal of scatter because single, rather than mean, $K_S$ observations were used. More recently \citet{Yuan2017b} discuss "mean" $JHK_S$ magnitudes for LMC Miras derived from a few observations combined with a variability model that comes from the much more extensive OGLE $I$-band observations. To these data they made linear and parabolic fits. Fig.~\ref{LMC_PL} shows the \citet{Yuan2017b} data together with 2MASS single observations for about 1600 Miras in the LMC; use of the mean photometry greatly reduces the scatter. 
The \citet{Ita2011} and \citet{Yuan2017b} fits are indistinguishable, so either could be used for distance scale measurements. The \citet{Whitelock2008} relation coincides with the other two for periods less than 400 days.

Although our understanding of the various PL sequences for SR variables has been improved recently, detailed modelling of the fundamental pulsators has not yet proved possible, even for the short period stars \citep{Trabucchi2017}, so it is unclear what shape of PL relation should really be anticipated. It seems likely that the break point, at around 400-420  days in the LMC, will be sensitive to the metallicity as it is due to the onset of HBB. The luminosity of the shorter period stars will be determined by the \citet{Paczynski1970} core-mass luminosity relation. The more luminous longer period stars are experiencing HBB at some level. Although there is no consensus on the details or even on the lower limit to mass of stars  which will experience HBB, it is expected to be abundance dependent \citep[][and references therein]{Ventura2015}. We should therefore predict that Gaia observations of Galactic O-rich Miras will show significant differences in the 400 to 500 day period range as HBB is expected only for higher mass stars in the higher metallicity environment.

\articlefiguretwo{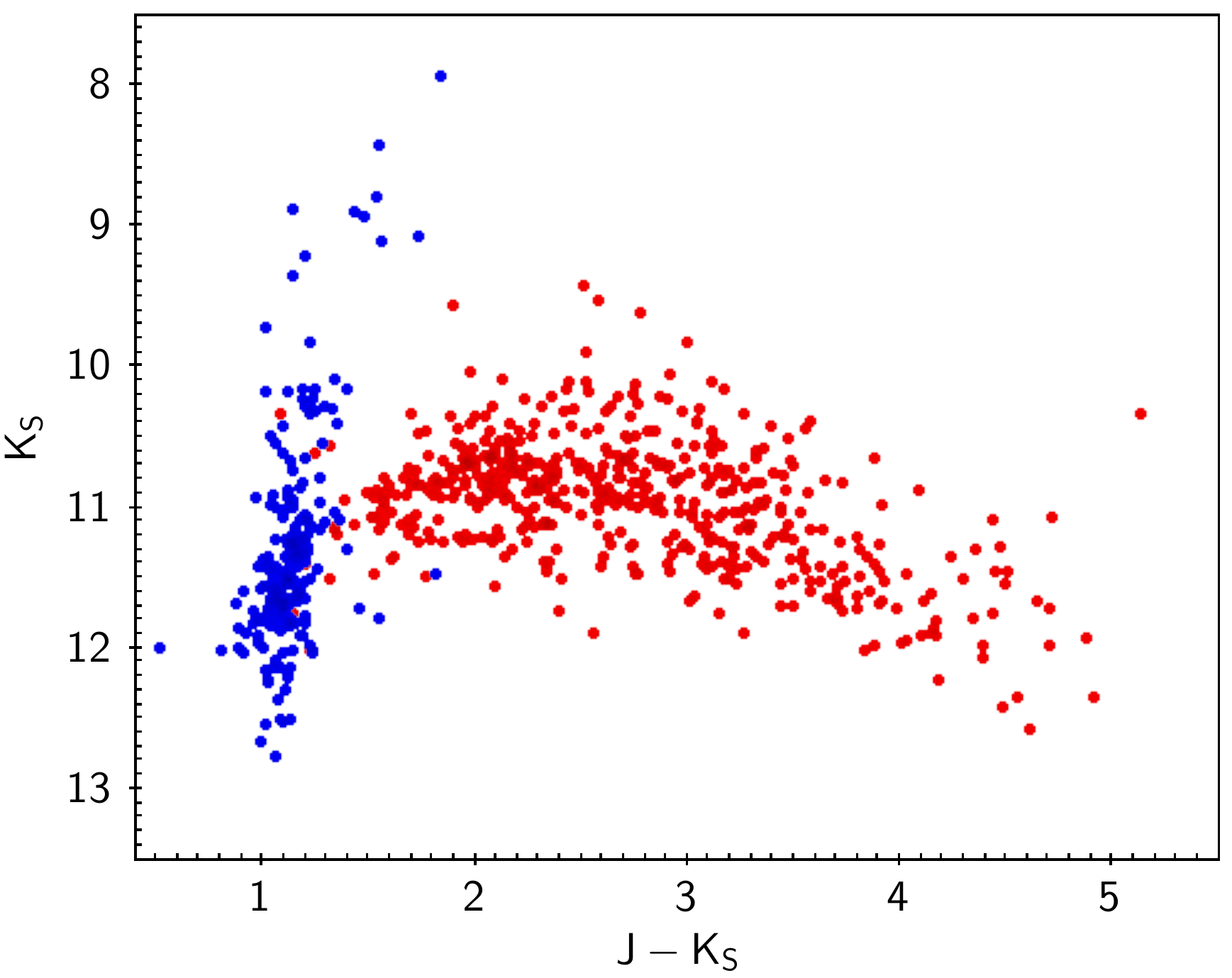}{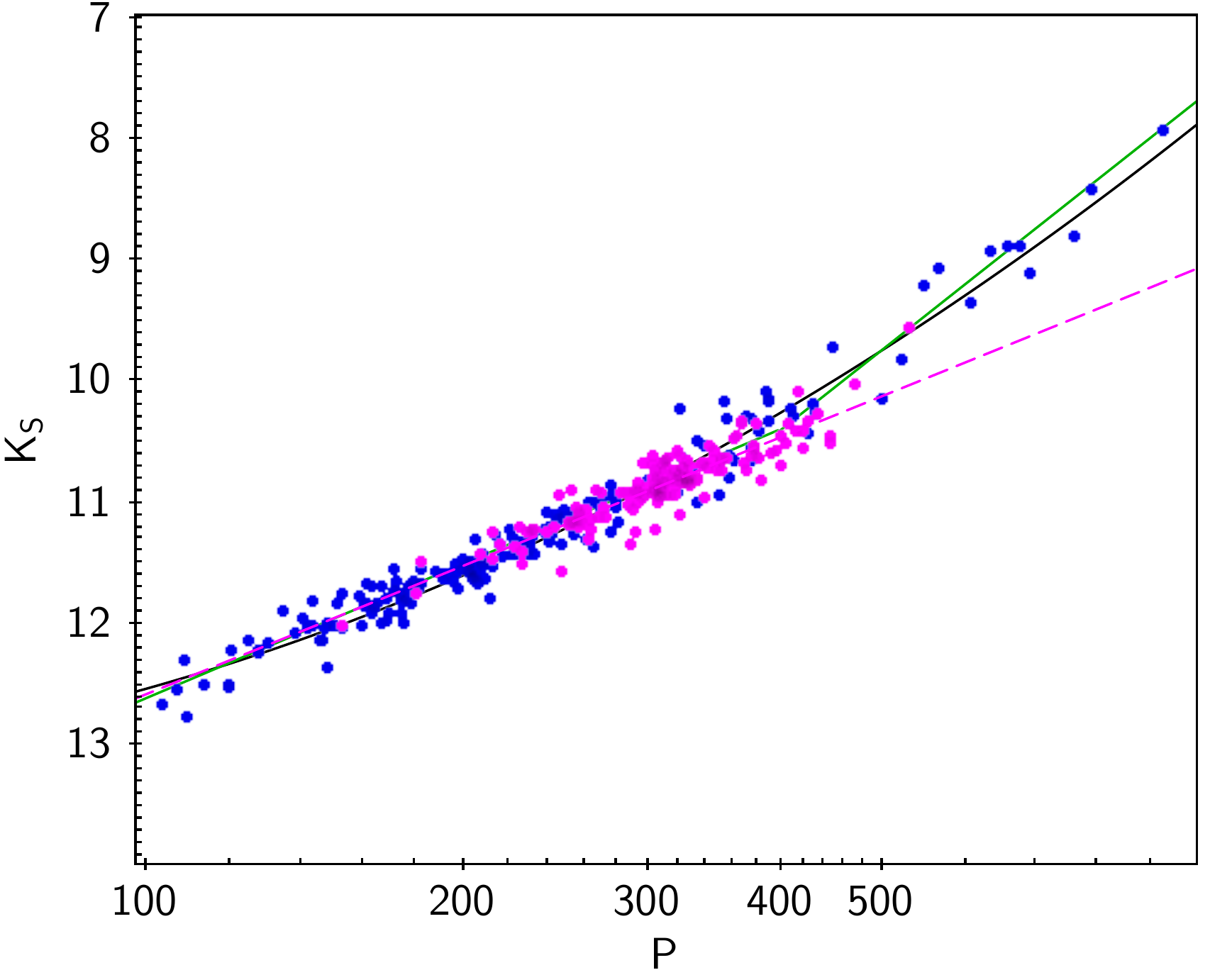}{LMC_PL2}{Miras for distance determination from the \cite{Yuan2017b} LMC sample: \emph{Left:} Colour magnitude diagram with O-rich and C-rich stars in blue and red, respectively; \emph{right:} PL relation for O-rich stars (blue) and C-rich stars with $J-K_S<2.2$ (magenta). The curves are the same as in Fig.~\ref{LMC_PL}.}

If one is primarily interested in Miras as standard candles for distance scale studies, then it is best to focus on the those with periods less than 400 days and with negligible circumstellar reddening\footnote{this may change as we develop a better understanding of the more luminous longer period stars}. The exact choice of waveband will depend on various factors, but $K$ has the advantage over shorter wavebands that the reddening (circumstellar and interstellar) is lower and the amplitude is smaller. It has the advantage over longer wavelengths that circumstellar emission is negligible; at [4.5] dust emission will contribute to the observed flux (see below). Fig.~\ref{LMC_PL2} shows that C-stars with $J-K_S<2.2$ follow that same $K_S$ PL relation that O-rich stars do. It has been known since \citet{Feast1989} that the $K$ PL  for short period Miras in the LMC was the same for C- and O-rich stars. This is despite the very different opacities and expectation that the radii and pulsation periods would be different for the two chemical types. It is therefore not essential to know the chemistry of the Miras, only their colour to know if they are reddened. Alternatively it is sometimes possible to correct for circumstellar extinction, as \citet{Ita2011} discuss. 

Although the short period stars are less luminous than their long period counterparts, the fact that they evolved from low mass progenitors means that they are more likely to be found in the outer regions of galaxies and therefore to be less crowded and more easily resolved at large distances.

Spitzer [3.6] and [4.5] observations of LMC AGB variables were discussed by \citet{Whitelock2017} and Fig.~\ref{fig} shows the [4.5] PL relation.   \citet{Riebel2015} derived PL relations for Mira variables, omitting the redder x-AGB stars, but this figure illustrates all of the Miras for which multiple observations and OGLE periods were available. The \citet{Riebel2015} PL relation for O-rich stars provides a good lower envelope for the O- and C-rich Miras and shows where most stars with very little dust emission lie. Stars with larger [3.6]--[4.5] colours are more luminous, and those where the dust shell dominates the [4.5] emission lie close to a PL relation that is about 1.5 mag brighter than the \citet{Riebel2015} relation. This is what we would expect if the star and dust shell can be approximated by blackbodies radiating at about 3000K and 300K, respectively.  It will be interesting  to see how stellar evolution models fit the details. The HBB stars do not stand out in this illustration, but are the triangles with $P\gtrsim420$ that lie between the two lines.

\articlefigure{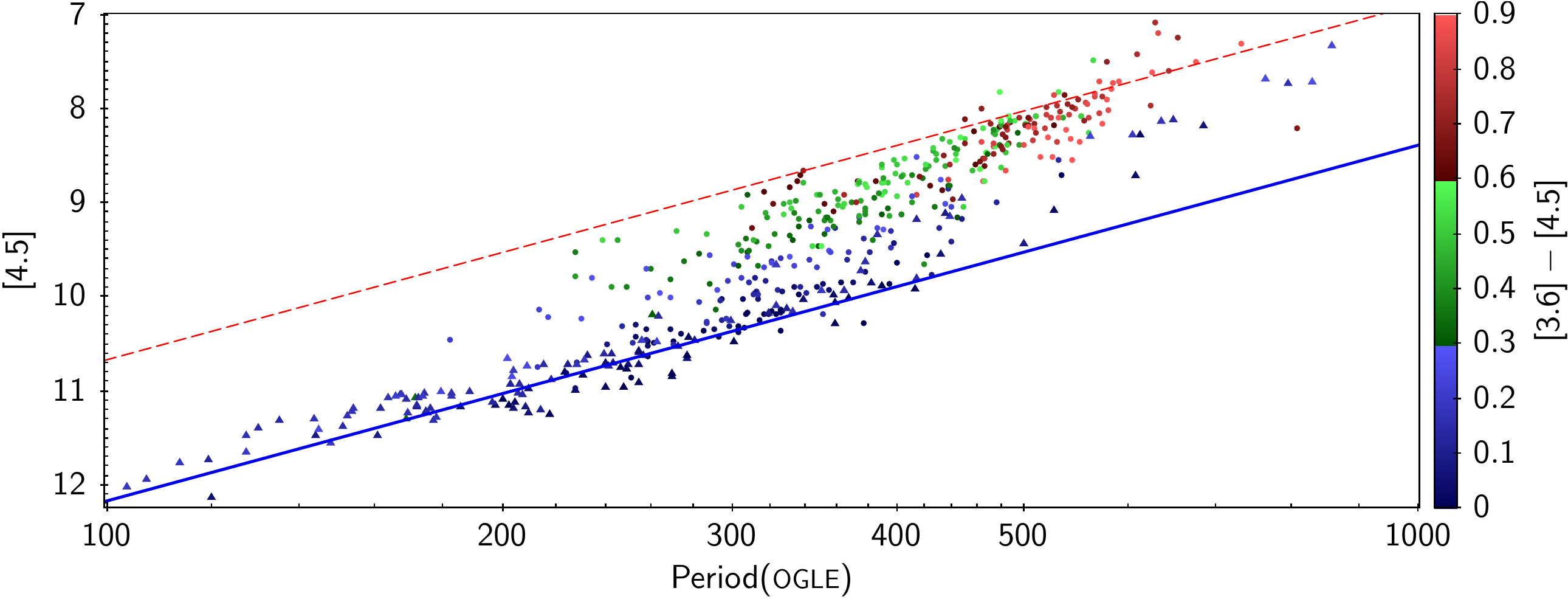}{fig}{The [4.5] PL relation for multiply observed LMC Miras. The photometry is from \citet{Riebel2015}, with O-rich and C-rich stars shown as triangles and circles respectively, colour coded according to [3.5]--[4.5]. The blue line is the  \citet{Riebel2015}  PL relation for O-rich Miras, the dashed red line is drawn parallel to the blue line, but 1.5 mag brighter.}

\section{NGC\,6822}
The AGB variables in NGC\,6822 were discussed by \citet{Whitelock2013} from a survey that was complete for HBB stars, which will be concentrated towards the centre of the galaxy, but would have missed C-stars that are either very red or in the periphery of the galaxy. Whitelock et al. discuss bolometric magnitudes, because the majority of the AGB variables are C-rich and have thick circumstellar shells; they derived these using a colour dependent bolometric correction. The results (Fig.~\ref{ngc6822}) show that the C-rich Miras are comparable to those in the LMC and distances derived from the Mira bolometric PL relation are very similar to those from other methods.

Note that these C-stars, whose periods range from  180 to 1000 days, fall on a linear PL, there is no sign of a steeper slope  for the longer period stars. This is consistent with the expectation that C-stars will not be HBB. Therefore if you want to correct the $K$ magnitude of a C-stars for circumstellar reddening you must  correct it to, e.g., the extrapolated \citet{Whitelock2008} relation rather than to one of those that fits the luminous HBB stars.

The four O-rich stars with periods between 540 and 640 days will probably be HBB, although the status of the slightly shorter period O-rich stars is unclear. The most luminous long period (P=854 days) O-rich star is probably a supergiant, although it is also a candidate super-AGB star and worth further investigation.

\articlefigure{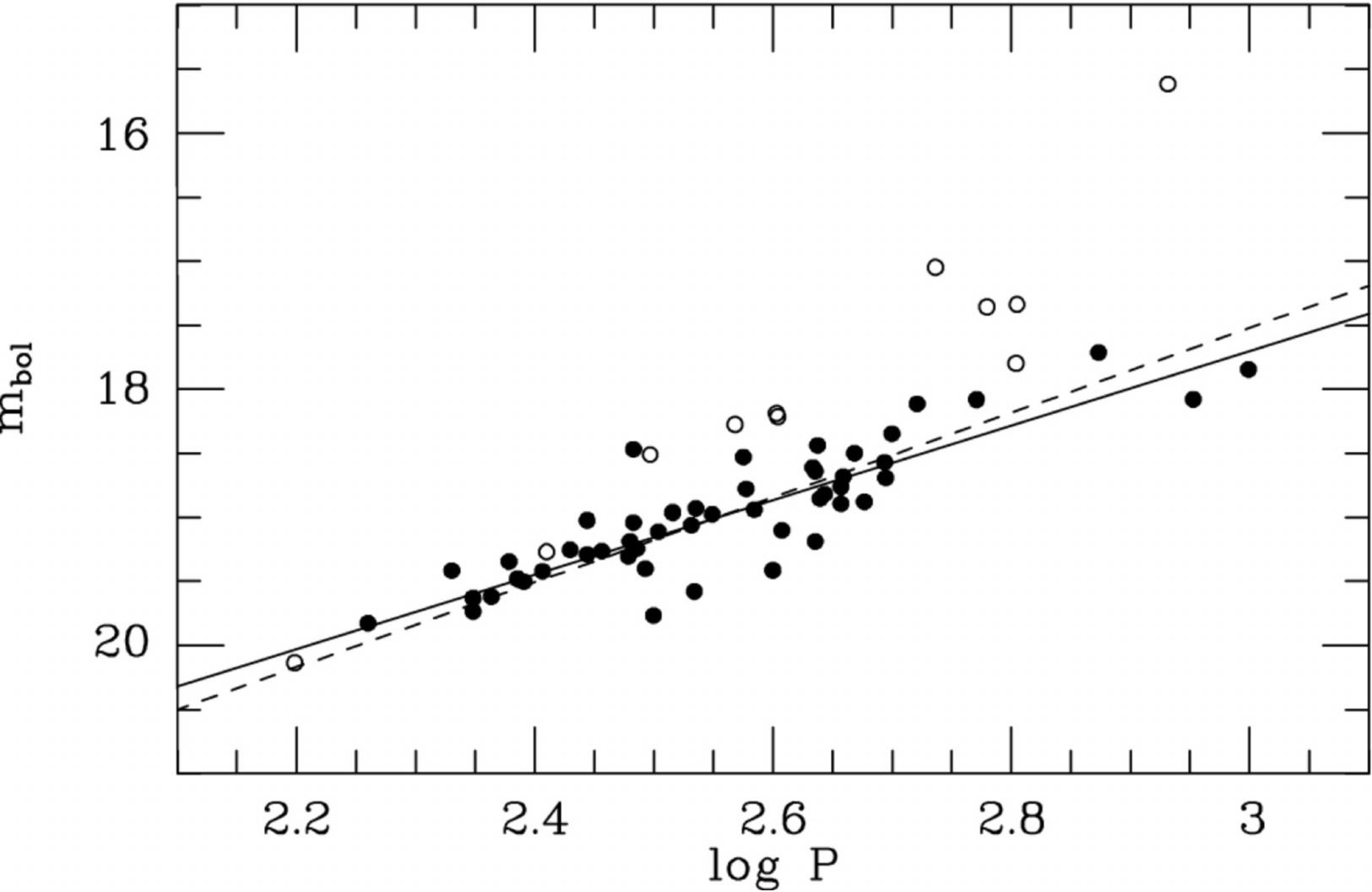}{ngc6822}{The bolometric PL relation for Miras in NGC\,6822 from \citet{Whitelock2013}; the open and closed circles represent O- and C-rich stars, respectively. The solid line is the PL relation derived from the C-rich stars in NGC\,6822 while the dashed line shows the best fitting relation with the same slope as the LMC PL relation for C-rich Miras.}

\section{IC\,1613}
The AGB variables in IC\,1613 were discussed by \citet{Menzies2015} and most notable was the identification of lithium in the spectrum of one of four HBB variables. The ages of these stars were estimated at about $2 \times 10^8$ yrs. The PL relation with C-rich Miras was used to estimate a distance  of 750\,kpc, in agreement with values in the literature. A preliminary analysis of Spitzer observations of these and other AGB variables was given by \citet{Whitelock2017}.

\section{Sagittarius Dwarf Irregular Galaxy (Sgr dIG)}
With $\rm  [Fe/H]=-1.88$ \citep{Kirby2017} Sgr dIG has the  lowest metallicity of any galaxy in which individual AGB variables have been studied. Three Mira variables were discovered by \citet{Whitelock2018}, two of them, V2 and V3, are C-rich with pulsation periods of 504 and 670 days and are very similar to the C-Miras in IC\,1613 and NGC\,6822.  The third, V1, has a period of 950 days, is O-rich \citep{Boyer2017}, bright and distinctly unusual. 

V1 is about a magnitude brighter at $K_S$ than the other AGB stars in Sgr~dIG so it really stands out in the colour-magnitude diagram. If such stars are present in more distant galaxies it should be possible to find them using large telescopes. Although it is bright at $K_S$ compared to most Miras, its position on the $K$ PL diagram  (Fig.~\ref{V1})  shows it to be at least one magnitude fainter than the PL relations obeyed by HBB stars and only slightly brighter than the shorter period HBB stars in IC\,1613. So it falls close to the extrapolated \citet{Whitelock2008} relation, where we find the reddening corrected carbon stars, which of course are not HBB. It is curious that there are no comparable stars in the LMC -- large amplitude variables with long periods, blue colours and luminosities close to the \citet{Whitelock2008} PL relation. This suggests that V1 is in a short lived evolutionary phase.

\articlefigure{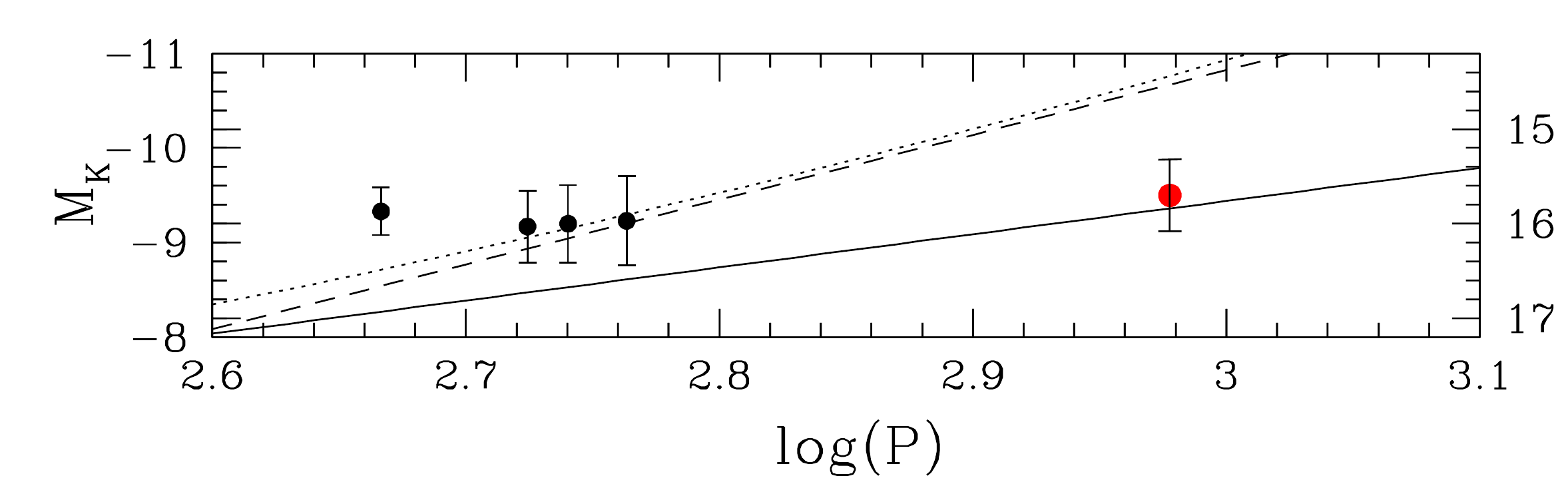}{V1}{Sgr dIG V1 (red) the O-rich Mira compared with the IC\,1613 HBB Stars (black) in a $K$ PL diagram. The dotted, dashed and solid lines are the PL relations from \citet{Yuan2017b}, \citet{Ita2011} and \citet{Whitelock2008}, respectively. The "error bars" show the full amplitude of the variability.} 

A comparison with stellar evolution models \citep{Marigo2013, Marigo2017} implies that the two carbon Miras, V2 and V3, had an initial mass around $3 M_{\odot}$. V1 would have been around $5 M_{\odot}$ and is now in a short lived and very advanced evolutionary state having terminated HBB. It will leave the AGB soon and we might anticipate  observable changes on a human timescale. 


\section{Conclusions}

There is still a great deal to learn about AGB variables, particularly the long period Miras, and the next generation of telescopes have a tremendous opportunity to investigate the short lived phases towards the end of the lives of intermediate mass stars. The short period Miras, which are better understood, have great potential as standard candles. 

\acknowledgements I am grateful to Lucas Macri for providing his data from \citet{Yuan2017b} in advance of full publication. Thanks to John Menzies and Michael Feast for their comments on a draft version of this paper. A grant from the South African National Research Foundation allowed me to travel to this meeting and I thank Richard de Gris for waiving the registration fee. I used TOPCAT \citep{Taylor2005} extensively in making this presentation.
\bibliography{pawbib.bib}

\begin{thebibliography}{}
\makeatletter
\relax
\def\mn@urlcharsother{\let\do\@makeother \do\$\do\&\do\#\do\^\do\_\do\%\do\~}
\def\mn@doi{\begingroup\mn@urlcharsother \@ifnextchar [ {\mn@doi@}
  {\mn@doi@[]}}
\def\mn@doi@[#1]#2{\def\@tempa{#1}\ifx\@tempa\@empty \href
  {http://dx.doi.org/#2} {doi:#2}\else \href {http://dx.doi.org/#2} {#1}\fi
  \endgroup}
\def\mn@eprint#1#2{\mn@eprint@#1:#2::\@nil}
\def\mn@eprint@arXiv#1{\href {http://arxiv.org/abs/#1} {{\tt arXiv:#1}}}
\def\mn@eprint@dblp#1{\href {http://dblp.uni-trier.de/rec/bibtex/#1.xml}
  {dblp:#1}}
\def\mn@eprint@#1:#2:#3:#4\@nil{\def\@tempa {#1}\def\@tempb {#2}\def\@tempc
  {#3}\ifx \@tempc \@empty \let \@tempc \@tempb \let \@tempb \@tempa \fi \ifx
  \@tempb \@empty \def\@tempb {arXiv}\fi \@ifundefined
  {mn@eprint@\@tempb}{\@tempb:\@tempc}{\expandafter \expandafter \csname
  mn@eprint@\@tempb\endcsname \expandafter{\@tempc}}}

\bibitem[\protect\citeauthoryear{{Boyer} et~al.,}{{Boyer}
  et~al.}{2017}]{Boyer2017}
{Boyer} M.~L.,  et~al., 2017, ApJ in press (arXiv:1711.02129), \href
  {http://adsabs.harvard.edu/abs/2017arXiv171102129B} {}

\bibitem[\protect\citeauthoryear{{Carpenter}}{{Carpenter}}{2001}]{Carpenter2001}
{Carpenter} J.~M.,  2001, \mn@doi [\aj] {10.1086/320383}, \href
  {http://adsabs.harvard.edu/abs/2001AJ....121.2851C} {121, 2851}

\bibitem[\protect\citeauthoryear{{Carter}}{{Carter}}{1990}]{Carter1990}
{Carter} B.~S.,  1990, \mn@doi [\mnras] {10.1093/mnras/242.1.1}, \href
  {http://ads.idia.ac.za/abs/1990MNRAS.242....1C} {242, 1}

\bibitem[\protect\citeauthoryear{{Feast}}{{Feast}}{2009}]{Feast2009}
{Feast} M.~W.,  2009, in {Ueta} T.,  {Matsunaga} N.,   {Ita} Y.,  eds, AGB
  Stars and Related Phenomena. p.~48 (\mn@eprint {arXiv} {0812.0250})

\bibitem[\protect\citeauthoryear{{Feast}, {Glass}, {Whitelock}  \&
  {Catchpole}}{{Feast} et~al.}{1989}]{Feast1989}
{Feast} M.~W.,  {Glass} I.~S.,  {Whitelock} P.~A.,   {Catchpole} R.~M.,  1989,
  \mn@doi [\mnras] {10.1093/mnras/241.3.375}, \href
  {http://adsabs.harvard.edu/abs/1989MNRAS.241..375F} {241, 375}

\bibitem[\protect\citeauthoryear{{Hughes} \& {Wood}}{{Hughes} \&
  {Wood}}{1990}]{Hughes1990}
{Hughes} S.~M.~G.,  {Wood} P.~R.,  1990, \mn@doi [\aj] {10.1086/115374}, \href
  {http://adsabs.harvard.edu/abs/1990AJ.....99..784H} {99, 784}

\bibitem[\protect\citeauthoryear{{Ita} \& {Matsunaga}}{{Ita} \&
  {Matsunaga}}{2011}]{Ita2011}
{Ita} Y.,  {Matsunaga} N.,  2011, \mn@doi [\mnras]
  {10.1111/j.1365-2966.2010.18056.x}, \href
  {http://adsabs.harvard.edu/abs/2011MNRAS.412.2345I} {412, 2345}

\bibitem[\protect\citeauthoryear{{Ita} et~al.,}{{Ita} et~al.}{2004}]{Ita2004}
{Ita} Y.,  et~al., 2004, \mn@doi [\mnras] {10.1111/j.1365-2966.2004.08126.x},
  \href {http://adsabs.harvard.edu/abs/2004MNRAS.353..705I} {353, 705}

\bibitem[\protect\citeauthoryear{{Kirby}, {Rizzi}, {Held}, {Cohen}, {Cole},
  {Manning}, {Skillman}  \& {Weisz}}{{Kirby} et~al.}{2017}]{Kirby2017}
{Kirby} E.~N.,  {Rizzi} L.,  {Held} E.~V.,  {Cohen} J.~G.,  {Cole} A.~A.,
  {Manning} E.~M.,  {Skillman} E.~D.,   {Weisz} D.~R.,  2017, \mn@doi [\apj]
  {10.3847/1538-4357/834/1/9}, \href
  {http://adsabs.harvard.edu/abs/2017ApJ...834....9K} {834, 9}

\bibitem[\protect\citeauthoryear{{Marigo}, {Bressan}, {Nanni}, {Girardi}  \&
  {Pumo}}{{Marigo} et~al.}{2013}]{Marigo2013}
{Marigo} P.,  {Bressan} A.,  {Nanni} A.,  {Girardi} L.,   {Pumo} M.~L.,  2013,
  \mn@doi [\mnras] {10.1093/mnras/stt1034}, \href
  {http://adsabs.harvard.edu/abs/2013MNRAS.434..488M} {434, 488}

\bibitem[\protect\citeauthoryear{{Marigo} et~al.,}{{Marigo}
  et~al.}{2017}]{Marigo2017}
{Marigo} P.,  et~al., 2017, \mn@doi [\apj] {10.3847/1538-4357/835/1/77}, \href
  {http://adsabs.harvard.edu/abs/2017ApJ...835...77M} {835, 77}

\bibitem[\protect\citeauthoryear{{Menzies}, {Whitelock}  \& {Feast}}{{Menzies}
  et~al.}{2015}]{Menzies2015}
{Menzies} J.~W.,  {Whitelock} P.~A.,   {Feast} M.~W.,  2015, \mn@doi [\mnras]
  {10.1093/mnras/stv1310}, \href
  {http://adsabs.harvard.edu/abs/2015MNRAS.452..910M} {452, 910}

\bibitem[\protect\citeauthoryear{{Paczy{\'n}ski}}{{Paczy{\'n}ski}}{1970}]{Paczynski1970}
{Paczy{\'n}ski} B.,  1970, Acta. Aston., \href
  {http://adsabs.harvard.edu/abs/1970AcA....20...47P} {20, 47}

\bibitem[\protect\citeauthoryear{{Riebel} et~al.,}{{Riebel}
  et~al.}{2015}]{Riebel2015}
{Riebel} D.,  et~al., 2015, \mn@doi [\apj] {10.1088/0004-637X/807/1/1}, \href
  {http://adsabs.harvard.edu/abs/2015ApJ...807....1R} {807, 1}

\bibitem[\protect\citeauthoryear{{Skrutskie} et~al.,}{{Skrutskie}
  et~al.}{2006}]{Skrutskie2006}
{Skrutskie} M.~F.,  et~al., 2006, \mn@doi [\aj] {10.1086/498708}, \href
  {http://ads.idia.ac.za/abs/2006AJ....131.1163S} {131, 1163}

\bibitem[\protect\citeauthoryear{{Soszy{\'n}ski} et~al.,}{{Soszy{\'n}ski}
  et~al.}{2009}]{Soszynski2009}
{Soszy{\'n}ski} I.,  et~al., 2009, Acta. Aston., \href
  {http://adsabs.harvard.edu/abs/2009AcA....59..239S} {59, 239}

\bibitem[\protect\citeauthoryear{{Taylor}}{{Taylor}}{2005}]{Taylor2005}
{Taylor} M.~B.,  2005, in {Shopbell} P.,  {Britton} M.,   {Ebert} R.,  eds,
  Astronomical Society of the Pacific Conference Series Vol. 347, Astronomical
  Data Analysis Software and Systems XIV. p.~29

\bibitem[\protect\citeauthoryear{{Trabucchi}, {Wood}, {Montalb{\'a}n},
  {Marigo}, {Pastorelli}  \& {Girardi}}{{Trabucchi}
  et~al.}{2017}]{Trabucchi2017}
{Trabucchi} M.,  {Wood} P.~R.,  {Montalb{\'a}n} J.,  {Marigo} P.,  {Pastorelli}
  G.,   {Girardi} L.,  2017, \mn@doi [\apj] {10.3847/1538-4357/aa8998}, \href
  {http://adsabs.harvard.edu/abs/2017ApJ...847..139T} {847, 139}

\bibitem[\protect\citeauthoryear{{Ventura}, {Karakas}, {Dell'Agli}, {Boyer},
  {Garc{\'{\i}}a-Hern{\'a}ndez}, {Di Criscienzo}  \& {Schneider}}{{Ventura}
  et~al.}{2015}]{Ventura2015}
{Ventura} P.,  {Karakas} A.~I.,  {Dell'Agli} F.,  {Boyer} M.~L.,
  {Garc{\'{\i}}a-Hern{\'a}ndez} D.~A.,  {Di Criscienzo} M.,   {Schneider} R.,
  2015, \mn@doi [\mnras] {10.1093/mnras/stv918}, \href
  {http://adsabs.harvard.edu/abs/2015MNRAS.450.3181V} {450, 3181}

\bibitem[\protect\citeauthoryear{{Whitelock} \& {Feast}}{{Whitelock} \&
  {Feast}}{2014}]{Whitelock2014}
{Whitelock} P.~A.,  {Feast} M.~W.,  2014, in EAS Publications Series. pp
  263--269, \mn@doi{10.1051/eas/1567047}

\bibitem[\protect\citeauthoryear{{Whitelock}, {Feast}, {van Loon}  \&
  {Zijlstra}}{{Whitelock} et~al.}{2003}]{Whitelock2003}
{Whitelock} P.~A.,  {Feast} M.~W.,  {van Loon} J.~T.,   {Zijlstra} A.~A.,
  2003, \mn@doi [\mnras] {10.1046/j.1365-8711.2003.06514.x}, \href
  {http://adsabs.harvard.edu/abs/2003MNRAS.342...86W} {342, 86}

\bibitem[\protect\citeauthoryear{{Whitelock}, {Feast}  \& {van
  Leeuwen}}{{Whitelock} et~al.}{2008}]{Whitelock2008}
{Whitelock} P.~A.,  {Feast} M.~W.,   {van Leeuwen} F.,  2008, \mn@doi [\mnras]
  {10.1111/j.1365-2966.2008.13032.x}, \href
  {http://adsabs.harvard.edu/abs/2008MNRAS.386..313W} {386, 313}

\bibitem[\protect\citeauthoryear{{Whitelock}, {Menzies}, {Feast}, {Nsengiyumva}
   \& {Matsunaga}}{{Whitelock} et~al.}{2013}]{Whitelock2013}
{Whitelock} P.~A.,  {Menzies} J.~W.,  {Feast} M.~W.,  {Nsengiyumva} F.,
  {Matsunaga} N.,  2013, \mn@doi [\mnras] {10.1093/mnras/sts188}, \href
  {http://adsabs.harvard.edu/abs/2013MNRAS.428.2216W} {428, 2216}

\bibitem[\protect\citeauthoryear{{Whitelock}, {Boyer}, {H{\"o}fner},
  {Wittkowski}  \& {Zijlstra}}{{Whitelock} et~al.}{2016}]{Whitelock2016}
{Whitelock} P.~A.,  {Boyer} M.,  {H{\"o}fner} S.,  {Wittkowski} M.,
  {Zijlstra} A.~A.,  2016, in 19th Cambridge Workshop on Cool Stars, Stellar
  Systems, and the Sun (CS19). p.~5 (\mn@eprint {arXiv} {1609.07954}),
  \mn@doi{10.5281/zenodo.154075}

\bibitem[\protect\citeauthoryear{{Whitelock}, {Kasliwal}  \&
  {Boyer}}{{Whitelock} et~al.}{2017}]{Whitelock2017}
{Whitelock} P.~A.,  {Kasliwal} M.,   {Boyer} M.,  2017, in {Catalan} M.,
  {Gieren} W.,  eds,  European Physical Journal Web of Conferences Vol. 152,
  {Wide-Field Variability Surveys: A 21st Century Perspective}. p.~1009
  (\mn@eprint {arXiv} {1702.06797})

\bibitem[\protect\citeauthoryear{{Whitelock}, {Menzies}, {Feast}  \&
  {Marigo}}{{Whitelock} et~al.}{2018}]{Whitelock2018}
{Whitelock} P.~A.,  {Menzies} J.~W.,  {Feast} M.~W.,   {Marigo} P.,  2018,
  \mn@doi [\mnras] {10.1093/mnras/stx2275}, \href
  {http://adsabs.harvard.edu/abs/2018MNRAS.473..173W} {473, 173}

\bibitem[\protect\citeauthoryear{{Wood}}{{Wood}}{2015}]{Wood2015}
{Wood} P.~R.,  2015, \mn@doi [\mnras] {10.1093/mnras/stv289}, \href
  {http://adsabs.harvard.edu/abs/2015MNRAS.448.3829W} {448, 3829}

\bibitem[\protect\citeauthoryear{{Wood} et~al.,}{{Wood}
  et~al.}{1999}]{Wood1999}
{Wood} P.~R.,  et~al., 1999, in {Le Bertre} T.,  {Lebre} A.,   {Waelkens} C.,
  eds,  IAU Symposium Vol. 191, Asymptotic Giant Branch Stars. p.~151

\bibitem[\protect\citeauthoryear{{Yuan}, {Macri}, {He}, {Huang}, {Kanbur}  \&
  {Ngeow}}{{Yuan} et~al.}{2017}]{Yuan2017b}
{Yuan} W.,  {Macri} L.~M.,  {He} S.,  {Huang} J.~Z.,  {Kanbur} S.~M.,   {Ngeow}
  C.-C.,  2017, \mn@doi [\aj] {10.3847/1538-3881/aa86f1}, \href
  {http://adsabs.harvard.edu/abs/2017AJ....154..149Y} {154, 149}

\makeatother
\end{thebibliography}

\end{document}